\documentstyle[12pt,epsf]{article}
\begin{document}
\baselineskip 16pt plus 2pt minus 2pt

\begin{titlepage}

\par
\topmargin=-1cm      

{ \small

\noindent{KFA-IKP(TH)-1997-16} \hfill{U.MDPP\# 98-009}

\noindent{DOE/ER/40762-127} \hfill{nucl-th/9708035}

}

\vspace{30.0pt}

\begin{centering}

{\large\bf{The isoscalar S-wave $\pi$-N scattering length
 $a^+$\\[0.3em]  from $\pi$-deuteron scattering}}

\vspace{40.0pt}
{{\bf S.R.~Beane}$^1$,
{\bf V.~Bernard}$^2$,
{\bf T.-S.H.~Lee}$^3$ and {\bf Ulf-G.~Mei{\ss}ner}$^4$}\\
\vspace{20.0pt}
{\sl $^{1}$Department of Physics, University of Maryland,
College Park, MD 20742, USA}\\
{\it E-mail address: sbeane@fermi.umd.edu}\\  
\vspace{15.0pt}
{\sl $^{2}$Laboratoire de Physique Th\'eorique} \\
{\sl Universit\'e Louis Pasteur, F-67037 Strasbourg Cedex 2, France} \\
{\it E-mail address: bernard@sbghp4.in2p3.fr}\\
\vspace{15.0pt}
{\sl $^{3}$Physics Division,
Argonne National Laboratory, Argonne, IL 60439, USA} \\
{\it E-mail address: lee@anlphy.phy.anl.gov}\\ 
\vspace{15.0pt}
{\sl $^{4}$Institut f\"ur Kernphysik, Forschungszentrum J\"ulich, 
D-52425 J\"ulich,
Germany} \\
{\it E-mail address: Ulf-G.Meissner@fz-juelich.de}\\
\vspace{15.0pt}
\end{centering}
\vspace{20.0pt}

\begin{abstract}
\noindent We consider constraints on the isoscalar 
S-wave $\pi$-N scattering
length $a^+$ from $\pi$-deuteron scattering, to third order in small
momenta and pion masses in chiral perturbation theory. To this order,
the $\pi$-deuteron scattering length is determined by $a^+$ together
with three-body corrections that involve no undetermined parameters.
We extract a novel value for a combination of dimension two
low--energy constants which is in agreement with previous determinations.
\end{abstract}

\vspace*{10pt}
\begin{center}
\end{center}
\vfill

\end{titlepage}


\noindent Chiral perturbation theory allows one to relate distinct scattering
processes in a systematic manner. Recently methodology has been
developed which relates scattering processes involving a single
nucleon to nuclear scattering processes~\cite{wein1}.  For instance,
one can relate $\pi$-N scattering to $\pi$-nucleus scattering. The
non-perturbative effects responsible for nuclear binding are accounted
for using phenomenological nuclear wavefunctions. Although this
clearly introduces an inevitable model dependence, one can compute
matrix elements using a variety of wavefunctions in order to ascertain
the theoretical error induced by the off-shell behavior of different
wavefunctions.

\noindent Weinberg showed that to third order ($O({q^3})$, where $q$ denotes a
small momentum or a pion mass) in chiral perturbation
theory the $\pi$-d scattering length is given by~\cite{wein1}

\begin{equation}
a_{\pi d}=\frac{(1+\mu)}{(1+\mu /2)}(a_{\pi n} + a_{\pi p})+{a^{(1b)}}+
{a^{(1c,1d)}},
\end{equation}
where $\mu\equiv{M_\pi}/m$ is the ratio of the pion and the nucleon
mass. The various diagrammatic contributions to
$a_{\pi d}$ are illustrated in figure~1. The three-body corrections
are (in momentum space):
\begin{equation}
{a^{(1b)}}= - \frac{{M_\pi^2}}{32{\pi^4}{f_\pi^4}{(1+\mu /2)}}
\langle\frac{1}{{\vec q}^{\,2}}\rangle_{\sl wf}
\end{equation}
\begin{equation}
{a^{(1c,1d)}}=\frac{{g_A^2}{M_\pi^2}}
{128{\pi^4}{f_\pi^4}{(1+\mu /2)}}
\langle\frac{{\vec q}\cdot{{\vec\sigma}_1}{\vec q}\cdot{{\vec\sigma}_2}}
{({\vec q}^{\,2}+{M_\pi^2})^2}\rangle_{\sl wf}.
\end{equation}
$\langle\vartheta\rangle_{\sl wf}$ indicates that 
$\vartheta$ is sandwiched
between deuteron wavefunctions.  These matrix elements have been
evaluated using a cornucopia of wavefunctions; results are in table 1.
Clearly ${a^{(1b)}}$ dominates the three-body corrections. This is the
result of the shorter range nature of $a^{(1c,1d)}$ as can be
seen from the r--space expressions of Eqs.(2) and (3). It is important
to stress that the dominant three--body correction turns out to
be quite independent of the wavefunction used. This implies that
the chiral perturbation theory approach, which relies on the dominance
of the pion--exchange, is useful in this context.

\noindent The $\pi$-N scattering lengths have the decomposition
\begin{equation}
a_{\pi n} + a_{\pi p}=2 a^{+}=2(a_{1}+2a_{3})/3,
\end{equation}
where $a^+$ is the isoscalar S-wave scattering length, and $a_{1}$ and
$a_{3}$ are the isospin $1/2$ and $3/2$ contributions, respectively.
Weinberg took $a^{+}$ from experimental data and argued that
${a^{(1b)}}$, which dominates the three-body corrections, should be
accounted for with corrections to the vertices, which he estimated
using a simple model~\cite{eric}. He then found a result for $a_{\pi
d}$ in agreement with the then current experimental value~\cite{pid}.
Since Weinberg's paper, there is new experimental information about
both the $\pi$-N and $\pi$-d scattering lengths that is at
variance with the old data~\cite{chat}\cite{sigg}. Moreover, since Eq.(1)
is a perfectly sensible expression to $O(q^3)$ in chiral perturbation
theory, we choose to take it seriously by using realistic deuteron
wavefunctions to evaluate both Eq.(2) and Eq.(3)
in order to see what it reveals.

\noindent We can express Eq.(1) as
\begin{equation}
a^{+}=\frac{(1+{\mu /2})}{2(1+\mu )}\biggl\lbrace a_{\pi d}-
({a^{(1b)}}+ {a^{(1c,1d)}})\biggr\rbrace ,
\end{equation}
and use experimental information about $\pi$-d scattering to predict
$a^+$; the recent PSI-ETHZ pionic deuterium measurement~\cite{chat}
gives
\begin{equation}
a_{\pi d}=-0.0264 \pm 0.0011\,{M_\pi^{-1}}.
\end{equation}
For the three-body corrections, we can safely ignore $a^{(1c,1d)}$ and
take the average of the $a^{(1b)}$ values in table~1:
\begin{equation}
a^{(1b)}=-0.02 \,{M_\pi^{-1}}.
\end{equation}
We then find
\begin{equation}
a^{+}=-(3.0 \pm 0.5)\,\cdot\,{10^{-3}}{M_\pi^{-1}},
\end{equation}
which is not consistent with the Karlsruhe-Helsinki value~\cite{koch},
\begin{equation}
a^{+}=-(8.3 \pm 3.8)\,\cdot\,{10^{-3}}{M_\pi^{-1}},
\end{equation}
or the new PSI-ETHZ value deduced from the strong interaction shifts in
pionic hydrogen and deuterium, which is small and 
positive~\cite{sigg}:\footnote{Note that this result might still change
  a bit since a more sophisticated treatment of
  Doppler--broadening for the width of the hydrogen level has to be 
  performed. Also, the PSI--ETHZ group did not yet quote a value for 
  $a^+$. We rather used their figure combining the H and d results
  to get the band given.}
\begin{equation}
a^{+}=(0...5)\,\cdot\,{10^{-3}}{M_\pi^{-1}}.
\end{equation}
The result Eq.(8) agrees, however, with the value obtained in the SM95
partial--wave analysis, $a^{+}=-3.0\,\cdot\,
{10^{-3}}{M_\pi^{-1}}$\cite{vpi}.
Given the ambiguous experimental situation regarding $a^{+}$, it seems
most profitable to turn our formula around and use the $\pi$-d
scattering data and three-body corrections to constrain undetermined
parameters that appear in $a^{+}$, which has been calculated to
$O({q^3})$ in chiral perturbation theory~\cite{bkm1}:
\begin{equation}
4\pi(1+\mu )a^{+}=
\frac{M_\pi^2}{F_\pi^2}\biggl(-4c_1 +2c_2 -\frac{g_A^2}{4m}+2c_3 \biggr)
+\frac{3{g_A^2}{M_\pi^3}}{64\pi{F_\pi^4}}.
\end{equation}
It should be stressed, however, that to this order there appear large
cancellations between the individual terms~\cite{bkm1} which lead one
to suspect that a calculation at $O({q^4})$ should be performed to
obtain a more precise prediction for this anomalously small
observable. This, however, goes beyond the scope of this manuscript. 
The sole undetermined parameter entering the $O({q^3})$ computation
of $a_{\pi d}$ is therefore a combination of $c_1$, $c_2$ and $c_3$: 
\begin{equation}
\Delta\equiv {-4c_1 +2(c_2 +c_3)}
\end{equation}
where we can now write
\begin{equation}
a_{\pi d}=\frac{1}{2\pi (1+\mu /2)}\biggl\lbrace 
\frac{M_\pi^2}{F_\pi^2}(\Delta -\frac{g_A^2}{4m})+
\frac{3{g_A^2}{M_\pi^3}}{64\pi{F_\pi^4}}\biggr\rbrace +
{a^{(1b)}}+ {a^{(1c,1d)}},
\end{equation}
and solve for $\Delta$:
\begin{equation}
\Delta =
\frac{2\pi{F_\pi^2}}{M_\pi^2}(1+\mu /2)
\lbrace a_{\pi d}-({a^{(1b)}}+{a^{(1c,1d)})}\rbrace
+\frac{g_A^2}{4m}\bigl(1-\frac{3m{M_\pi}}{16\pi{F_\pi^2}}\bigr)
\end{equation}
in order to constrain $\Delta$ using Eqs.(2), (3) and (6).  We find
\begin{equation}
\Delta =-(0.10\pm 0.03)\, {\rm GeV}^{-1},
\end{equation}
where we have taken into account the error in the determination of
$a_{\pi d}$.

\noindent In table~2 we give values of the relevant $c_i$'s obtained from a
realistic fit to low-energy pion-nucleon scattering
data and subthreshold parameters~\cite{bkm2}. 
Central values lead to $\sigma (0)=47.6\,$MeV and
$a^+ =-4.7\cdot 10^{-3}{M_\pi^{-1}}$. These values of the 
$c_i$'s give the conservative determination:
\begin{equation}
\Delta =-(0.18\pm 0.75)\, {\rm GeV}^{-1}.
\end{equation}
Also shown in table~2 are values of $c_i$'s deduced from resonance
saturation. It is worth mentioning that an independent fit to
pion-nucleon scattering including also low--energy constants related
to dimension three operators finds results consistent with the fit values
of table~2~\cite{moj}.

\noindent To summarize, we have shown that the recent precise data on
the $\pi$--deuteron scattering length can be used to constrain a
combination of dimension two low--energy constants of the chiral
effective pion--nucleon Lagrangian. This determination gives a result
in agreement with previous determinations that use independent
input~\cite{bkm2}\cite{moj}. Therefore, a consistent picture of
nucleon chiral perturbation theory is emerging. Next, these
calculations should be carried out one order further which would allow
one to {\it precisely} deduce the isoscalar S--wave $\pi$-N scattering
length from the accurately measured $\pi$-d scattering length. Work
along these lines is in progress.

\bigskip\bigskip

\noindent{\bf Acknowledgments:}

\smallskip

\noindent VB and UGM are grateful to the Nuclear Theory Group at
Argonne National Laboratory for hospitality while part of this work
was completed. We thank R.~Workman for a useful communication.  This
research was supported in part by the U. S. Department of Energy,
Nuclear Physics Division (grants DE-FG02-93ER-40762 (SRB),
W-31-109-ENG-38 (TSHL) ), by NATO Collaborative Research Grant 950607
(VB, TSHL, UGM) and by the Deutsche Forschungsgemeinschaft (grant ME
864/11-1 (UGM)).
\newpage
\begin{center}

\begin{tabular}{|l|r|r|r|c|}
    \hline ${\sl wf}$ & $a^{(1b)}$ & $a^{(1c,1d)}$ \\ \hline 
    Bonn\protect{\cite{bonn}} &
  $-0.02021$ & $-0.0005754$ \\ 
   ANL-V18\protect{\cite{v18}} & $-0.01960$ & $-0.0007919$ \\
   Reid-SC\protect{\cite{reid}} & 
  $-0.01941$ & $-0.0008499$ \\ 
   SSC\protect{\cite{ssc}} & 
  $-0.01920$ & $-0.0006987$ \\ \hline \end{tabular}

\smallskip 

\end{center}

\noindent Table 1: Three-body corrections for various
deuteron wavefunctions in units of $M_\pi^{-1}$.  We use
${F_\pi}=92.4$\,MeV, ${g_A}=1.32$ and ${M_{\pi^+}}=139.6\,$MeV.


\begin{center}

\begin{tabular}{|l|r|r|r|c|}
    \hline
    $i$         & $c_i \quad \quad$   &  
                  $ c_i^{\rm Res} \,\,$ cv & 
                  $ c_i^{\rm Res} \,\,$ ranges    \\
    \hline
    1  &  $-0.93 \pm 0.10$  &  $-0.9^*$ & -- \\
    2  &  $3.34  \pm 0.20$  &  $3.9\,\,$ & $2 \ldots 4$ \\    
    3  &  $-5.29 \pm 0.25$  &  $-5.3\,\,$ 
                                     & $-4.5 \ldots -5.3$ \\
    \hline
    $\Delta$ & $-0.18 \pm 0.75$ & $0.8 \,\, $& $-3.0 \ldots +2.6$\\ 
    \hline
  \end{tabular}

\smallskip 

\end{center}

\noindent Table 2: Values of the LECs $c_i$ in GeV$^{-1}$
for $i=1,\ldots,3$.  Also given are the central values (cv) and the
ranges for the $c_i$ from resonance exchange. The $^*$ denotes an
input quantity. This table is adopted from~\cite{bkm2}.

\begin{figure}[h]
   \vspace{0.5cm} \epsfysize=14cm
   \centerline{\epsffile{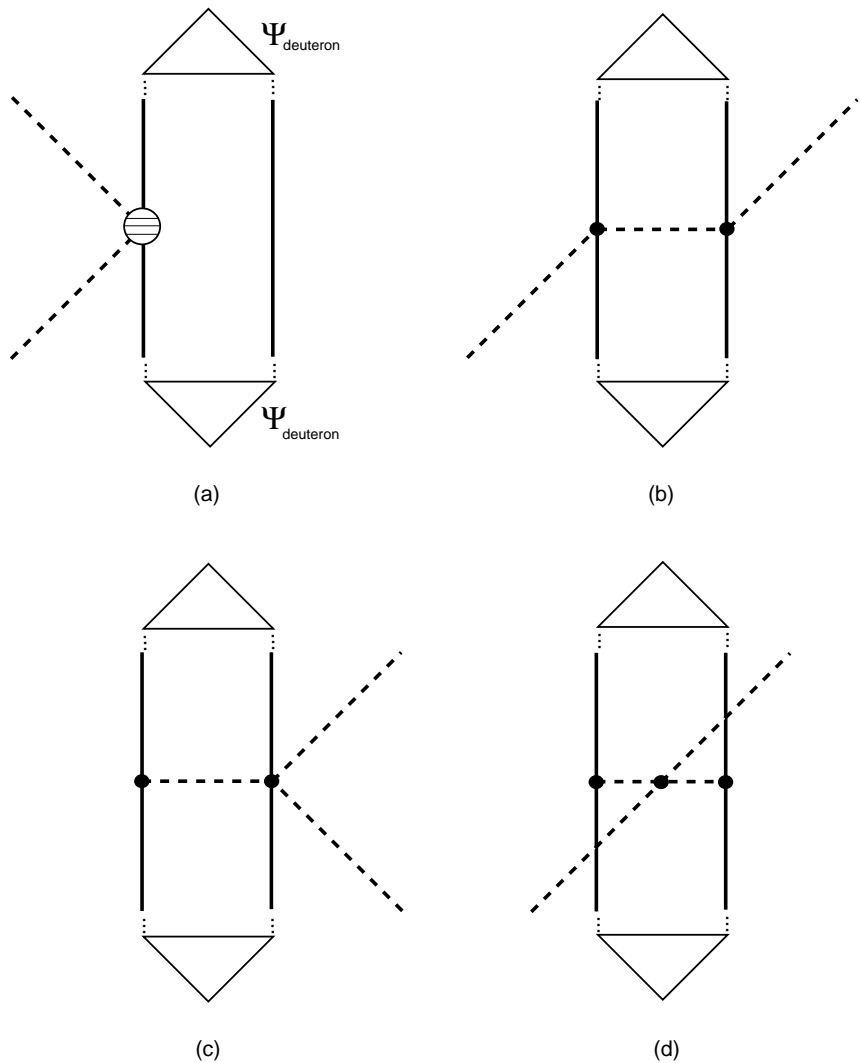}}
   \centerline{\parbox{11cm}{\caption{\label{fig1} Feynman graphs
   contributing to the $\pi$-d scattering length at order $q^3$ in
   chiral perturbation theory. Graph (a) is the single scattering
   contribution and contains undetermined parameters. Graphs (b), (c)
   and (d) are three--body interactions which involve no undetermined 
   parameters.
   }}}
\end{figure}


\end{document}